\documentclass[epj]{svjour}
\usepackage{graphics}
\begin{document}
\renewcommand{\thefootnote}{\alph{footnote}}
\title{Spin Glass and antiferromagnetism in Kondo-lattice disordered systems}
\author{S. G. Magalh\~aes\inst{1,\mbox{\scriptsize a}} 
\and A. A. Schmidt\inst{2,\mbox{\scriptsize b}}\and F.\ M.\ Zimmer\inst{1} 
\and A. Theumann\inst{3,\mbox{\scriptsize c}}
\and B. Coqblin\inst{4,\mbox{\scriptsize d}}}

%
\institute{ Departamento de F\'\i sica -- UFSM,  97105-900 Santa Maria, RS, Brazil 
\and Departamento de Matem\'atica -- UFSM,  97105-900 Santa Maria, RS, Brazil
\and Instituto de F\'\i sica -- UFRGS, 91501-970 Porto Alegre, RS, Brazil
\and Laboratoire de Physique des Solides Universit\'e Paris-Sud, 91405 Orsay, France}
\date{Received: date / Revised version: date}
%
\abstract{
The competition between spin glass (SG), 
antiferromagnetism (AF) and Kondo effect is studied here in a model which consists of two Kondo 
sublattices with a gaussian random  interaction between 
spins in differents sublattices  
with an antiferromagnetic mean $Jo$ and standard deviation $J$. 
In the present approach there is no hopping of the conduction electrons between the sublattices 
and only spins in different sublattices can interact. The problem is formulated in 
the path integral formalism where the spin operators are expressed as bilinear 
combinations of Grassmann fields which can be solved at mean field level within the 
static approximation and the replica symmetry ansatz. The obtained phase diagram 
shows the sequence of phases SG, AF and Kondo state for increasing Kondo 
coupling. This sequence agrees qualitatively with experimental data of 
the $Ce_{2} Au_{1-x} Co_{x} Si_{3}$ compound.
} 
\maketitle
\footnotetext[1]{\email{ggarcia@ccne.ufsm.br}}
\footnotetext[2]{\email{alex@lana.ccne.ufsm.br}}
\footnotetext[3]{\email{albath@if.ufrgs.br}}
\footnotetext[4]{\email{coqblin@lps.u-psud.fr}}

\section{Introduction}
\label{intro}
It is recognized that there is a strong competition between the Kondo effect and the RKKY interaction
in Kondo lattice systems \cite{Coqblin}. A transition from a magnetically ordered phase 
to a heavy fermion one, described by a Fermi-Liquid behaviour, has been observed in many $Ce$ 
or $Yb$ compounds and extensively studied from a theoretical point of view. 
There is a quantum critical point (QCP) at the transition and non-Fermi liquid behaviors are also observed 
near the QCP \cite{Coleman}. The role of disorder has been studied by different approaches including 
a "Kondo disorder" model describing a broad distribution of Kondo temperatures \cite{Miranda} or the extensive study of the so-called 
"quantum Griffiths" behaviour \cite{Castro}.  On the opposite, the transition from a metallic spin glass phase
to a paramagnetic or Kondo phase has been studied recently  by using the quantum rotor spin glass model \cite{Georges} 
and the existence of an anomalous behavior 
near $T=0$ has been observed
in the transition between a metallic paramagnetic and a metallic spin glass.

Particularly,  spin glass (SG) and Kondo state have been found together in some Cerium 
alloys like $Ce$$Ni_{1-x}$ $Cu_{x}$ \cite{Gomez-Sal}, 
$Ce_{2} Au_{1-x} Co_{x} Si_{3}$ \cite{Majundar} and in some disordered Uranium alloys such as 
$UCu_{5-x}Pd_{x}$ \cite{Vollmer} or $U_{1-x}La_{x}Pd_{2}Al_{3}$ \cite{Zapf}. 
In the first case, there is an 
antiferromagnetic (AF) phase for low contents of $Ni$. When the $Ni$ doping is 
increased the phase diagram becomes more complex. For $x<0.8$, the sequence 
of phases SG-ferromagnetism arises when the temperature is lowered 
and a Kondo state exists for $x<0.2$.  The alloys $Ce_{2}Au_{1-x}Co_{x}Si_{3}$ 
exhibit a phase diagram with a sequence of SG, AF and non magnetic Kondo phases with increasing 
the cobalt concentration at low temperature; in the high doping situation, the Neel 
temperature seems to tend to zero. In the two previously mentioned cases of disordered alloys 
\cite{Vollmer,Zapf}, the AF, SG and NFL phases have been obtained at low temperatures 
for different concentrations; in particular, the sequence AF-NFL-SG occurs with increasing $x$ 
in $UCu_{5-x}Pd_{x}$ alloys \cite{Vollmer} and the opposite sequence AF-SG-NFL with increasing $x$ 
in $U_{1-x}La_{x}Pd_{2}Al_{3}$ alloys \cite{Zapf}. 

Thus, the results previously mentioned evidence a quite complicated 
interplay among the Kondo effect and the RKKY interaction when disorder and 
frustration are pre\-sent. Recently, a theoretical effort has been done to 
understand how a spin glass phase emerges in a Kondo lattice model with an intrasite
exchange interaction and an intersite long range random interaction among the localized spins  
\cite{Alba}. The mentioned model has been extended to produce also a ferromagnetic order 
\cite{Magal}. A particularity of our model \cite{Alba,Magal} compared to those discussed in Ref. \cite{Georges} is that we do not
consider quantum fluctuations by coupling only the $S_{z}$ components of the localized 
spins in the spin glass term of the Hamiltonian, thus being unable to discuss a QCP at $T=0$. One important remaining issue would be to obtain a description of the interplay
between antiferromagnetism, Kondo phase and spin glass. The first step in that direction would be to produce a theory at mean 
field level able to mimic some important aspects, for instance, the sequence 
of magnetic phases of $Ce_2 Au_{1-x}Co_{x}Si_3$ in terms of a 
minimum set of energy scales related to the 
fundamental mechanisms present in the problem. In this work this set is $J_0$ 
(the random intersite average), $J$ (the random intersite variance) and $J_{K}$ 
(the strength of the Kondo coupling). In the following, we take a constant $J_{K}$ and a
Gaussian distribution of the intersite interaction with a random intersite interaction with average $J_{0}$ and variance $J$. 
Thus, we will not take values of $J_{K}$ and $T_{K}$ given by a random distribution
as in the work of Miranda et all \cite{Miranda}.

In view of that, the aim of this paper is to present a mean field theory to 
study the interplay among disorder and antiferromagnetic ordering in a 
Kondo lattice providing the necessary refinements in the previous theory 
\cite{Alba,Magal}. 
The model studied is a two Kondo sublattice with an intrasite exchange interaction 
and an intersite random 
gaussian interaction only between localized spins in two different sublattices 
\cite{Korenblit}. We consider here that there is no hopping of the conduction electrons between the two different sublattices; this assumption, which is used here to really simplify the calculations, does not modify the results since the different AF or SG ordering are in fact essentially due to the localized spins.   
The possible AF or SG ordering is, therefore, entirely related to the coupling of localized spins between the two sublattices.   
This fermionic problem is formulated by writing the spins 
operators as bilinear combinations of Grassmann fields and, through the 
static approximation and the replica 
formalism, the partition function is obtained, as already introduced to treat at mean field level the spin glass, Kondo 
effect and ferromagnetism in Refs. \cite{Alba} and \cite{Magal}.

The Kondo  effect and the RKKY interaction originate from the same intrasite 
exchange interaction, but the necessity of considering an additional intersite exchange 
term has been already recognized \cite{Ueda} and the full Hamiltonian with both terms
has been extensively used to describe the Kondo lattice \cite{Coqblin1,Iglesias}. Moreover, 
the presently studied model has explicitly a random coupling term among localized spins in order to 
describe the spin glass situation \cite{Alba,Magal}. The 
{\it exchange terms} cannot be considered as completely independent from each other and a relationship
giving the intersite exchange integral varying as the square of the Kondo local exchange 
integral $J_{K}$ has been introduced in order to mimic the $(J_{K})^{2}$ dependence of 
the RKKY interaction \cite{Iglesias}. In fact the relationship corresponds to an approximate representation
of the intersite interaction, but we will use it in the last section of our paper in order to have a
better agreement with experiment for some disordered 
Cerium alloys. Recently, the Doniach diagram has been revisited and the 
suppression of the AF and SG phases investigated, although these orders have been considered as
two independent problems \cite{Oppermann}.

This paper is structured as follows. In Section II the model is introduced and 
developed in order to get the free energy and the corresponding saddle point 
equations for the order parameters. In Section III, as mentioned in the previous paragraph,
a relationship among $J_0$, $J$ 
and $J_{K}$ is introduced allowing to solve the order parameter equations and to 
build up a temperature {\em versus} $J_{K}$ phase diagram showing the sequence of 
phases SG-AF-Kondo state. The conclusions are presented in the last section.

\section{General Formulation}
\label{sec:1}

The model considered here consists in two Kondo sublattices $A$ and $B$ with a random coupling only 
between localized spins in distinct sublattices \cite{Korenblit}. 
Furthermore, there is no hopping of conduction electrons between the sublattices as previously explained. 
The corresponding Hamiltonian is given by:
\begin{eqnarray} 
\displaystyle{\cal H}-\mu N=\sum_{p=A,B}[\sum_{i,j}\sum_{\sigma=\uparrow \downarrow}
 t_{i j}\hat{d}_{i,p,\sigma}^{\dagger}\hat{d}_{j,p,\sigma} +
\sum_{i}\varepsilon_{0, p}^{f}\hat{n}_{i,p}^{f}~~ \nonumber\\+
J_{K}(\sum_{i}\hat{S}_{i,p}^{+}\hat{s}_{i,p}^{-}+\hat{S}_{i,p}^{-}\hat{s}_{i,p}^{+})]+
\sum_{i j} J_{i j}\hat{S}_{i,A}^{z}\hat{S}_{j,B}^{z}~~
\label{e1}
\end{eqnarray}
\noindent where $i$ and $j$ sums run over $N$ sites of each sublattice. 
The intersite interaction $J_{i j}$ is assumed to be a random quantity 
following a gaussian distribution \cite{Korenblit}
\begin{eqnarray} 
\displaystyle {\cal P}{(J_{i j})}=\frac{1}{J}\sqrt{\frac{N}{64\pi}}
\exp\left\{-\frac{(J_{i j}+2J_{0}/N)^{2}}{64J^{2}}N\right\}.
\label{e2}
\end{eqnarray}
\noindent An anti-ferromagnetic solution for the present choice of $J_{i j}$ 
can be found for $J_{0} > 0$. The case $J_{0} < 0$ produces a complex phase 
diagram with spin glass, ferromagnetic, mixed phase (a spin glass with 
spontaneous magnetization) \cite{Magal} and a Kondo state as defined in the Ref.
\cite{Alba}.

The spin variables present in the Eq.\ (\ref{e1}) are 
de\-fi\-ned clo\-se\-ly to Ref.\ \cite{Alba} mo\-di\-fied for the case of 
two interacting distinct sub-lattices given as: 
$\displaystyle \hat{s}_{i,p}^{+}=\hat{d}_{i,p,\uparrow}^{\dagger}
\hat{d}_{i,p,\downarrow \uparrow}=(\hat{s}_{i,p}^{-})^{\dagger}
$, $\displaystyle \hat{S}_{i,p}^{+}=\hat{f}_{i,p,\uparrow}^{\dagger}
\hat{f}_{i,p,\downarrow }=(\hat{S}_{i,p}^{-})^{\dagger}
$ 
and 
$\displaystyle \hat{S}_{i,p}^{z}= \frac{1}{2} 
\left[\hat{f}_{i,p,\uparrow}^{\dagger}
\hat{f}_{i,p,\uparrow }\right.$ $-\hat{f}_{i,p,\downarrow}^{\dagger}$ 
$\left.\hat{f}_{i,p,\downarrow }\right]$ where $\hat{d}_{i,p, \sigma}^{\dagger}$, $\hat{d}_{i,p,\sigma}$ $(\hat{f}
_{i,p, \sigma}^{\dagger},\hat{f}_{i,p, \sigma})$ 
are the creation and destruction operators for conduction (localized) fermions. 

The partition function can be given by using Grassman variables 
$\psi_{i,p,\sigma}(\tau)$ for the localized fermion and 
$\varphi_{i,p,\sigma}(\tau)$ for the conducting ones:
\begin{eqnarray}
\displaystyle Z=\int\!\!\!\prod_{p=A,B}D(\psi_{i,p,\sigma}^{*}\psi_{i,p,\sigma})
\int\!\!\!\prod_{p=A,B}D(\varphi_{i,p,\sigma}^{*}\varphi_{i,p,\sigma})\,e^{A}
\label{e6}
\end{eqnarray}
\noindent where the action is $A= A_{0}+A_{K}+ A_{SG}$ with
\begin{eqnarray} 
 A_0=\sum_{p=A,B}\sum_{i,j}\sum_{\omega}
 \left(\gamma_{i j,p}^{0}(\omega)\right)^{-1}
\left[\varphi_{i,p,\uparrow}^{*}(\omega)\varphi_{j,p,\uparrow}(\omega) \right.
\nonumber\\+ \left.
\varphi_{i,p,\downarrow}^{*}(\omega)\varphi_{j,p,\downarrow}(\omega)\right]
\nonumber\\ + 
\sum_{p=A,B}\sum_{\omega}\sum_{i j}\left(g_{i j, p}^{0}(\omega)\right)^{-1}
 \left[\psi_{i,p,\uparrow}^{*}(\omega) \psi_{j,p,\uparrow}(\omega)\right. 
 \nonumber\\+ \left. 
 \psi_{i,p,\downarrow}^{*}(\omega) \psi_{j,p,\downarrow}(\omega) \right]
\label{e7}
\end{eqnarray}
\noindent and
\begin{eqnarray} 
\left[\gamma_{i j,p}^{0}(\omega)\right]^{-1}&=&(i\omega -\beta \varepsilon_{0,p}^{c})
\delta_{ij}-t_{i j} \\
\label{e8}
\left[g_{i j,p}^{0}(\omega)\right]^{-1}&=&(i\omega -\beta 
\varepsilon_{0, p}^{f})\delta_{i,j}~.
\label{e81}
\end{eqnarray}
\noindent The energies 
$\varepsilon_{0, A}^{c}=\varepsilon_{0, B}^{c}=\varepsilon_{0}^{c}$ 
and 
$\varepsilon_{0, A}^{f}=\varepsilon_{0, B}^{f}=\varepsilon_{0}^{f}$ 
are referred to the chemical potentials of the conduction and localized 
bands, respectively. 

The actions $A_{SG}$ and $A_{K}$ are given by
\begin{equation} 
A_{SG}=\beta\sum_{i j}\sum_{\nu}J_{i j} S_{i,A}^{z}(\nu) 
S_{j,B}^{z}(-\nu)
\label{e9}
\end{equation}
\noindent and
\begin{eqnarray} 
A_K=\beta\frac{J_K}{N}\sum_{\sigma=\uparrow\downarrow}\sum_{p=A,B}\left[\sum_{i=1}^{N}\sum_{\omega}
\varphi_{i,p,-\sigma}^{*}(\omega)\psi_{i,p,-\sigma}(\omega)\right]
\times\nonumber \\
~~~\left[\sum_{i^{'}=1}^{N}\sum_{\omega^{'}}
\psi_{i^{'},p,\sigma}^{*}(\omega^{'})\varphi_{i^{'},p,\sigma}(\omega^{'})\right]
~~~
\label{e10}
\end{eqnarray}
\noindent with Matsubara's frequencies $\omega=(2m+1)\pi$ and 
$\nu=2m\pi$ $(m=0,\pm 1, \pm 2,\ldots )$.

The static approximation will be used in Eq.\ (\ref{e9}) and Eq.\ (\ref{e10}) to
solve this problem at mean field level. In that spirit, the Kondo state at one
particular sub-lattice $s$ is caracterized by the complex order parameter
$\lambda_{p,\sigma}=\frac{1}{N}$ $\sum_{i,\omega}$ $\left<\varphi_{i,p,\sigma}^{*}
(\omega)\psi_{i,p,\sigma}(\omega)\right>$ ($p=A,B$) which is introduced in 
the present theory through the identity
\begin{eqnarray}
\delta\left( N\lambda_{p,\sigma}-\sum_{\omega}\sum_{i=1}^{N}
\varphi_{i,p,\sigma}^{*}(\omega)\psi_{i,p,\sigma}(\omega) \right)
=\int\prod_{\sigma}\frac{dv_{p,\sigma}}{2\pi}\times\nonumber\\
\exp\left\{i\sum_{\sigma}v_{p,\sigma}\left[
N\lambda_{p,\sigma}^{*}-\sum_{\omega}\sum_{i=1}^{N}
\varphi_{i,p,\sigma}^{*}(\omega)\psi_{i,p,\sigma}(\omega)\right]\right\}.~~~
\label{e12}
\end{eqnarray}
\noindent Its conjugate, $\lambda_{p,\sigma}^{*}$, can be also introduced 
by a similar identity. From now on, it is assumed that $\lambda_{p,\sigma}\approx\lambda_{p}$ 
($\lambda^{*}_{p,\sigma}\approx\lambda^{*}_{p}$) \cite{Alba,Magal}.
 
Therefore, the partition function given in Eqs.\ (\ref{e6})-(\ref{e10}), after using the 
integral representation for the delta functions, is
\begin{equation}
Z=exp\{-2N\beta J_k(|\lambda_A|^2+|\lambda_B|^2)\}\,Z^{(stat)}
\label{e121}
\end{equation}
\noindent where
\begin{eqnarray}
 Z^{(stat)}\!\!&=&\!\!\int\!\!\prod_{p=A,B}{\cal D}(\psi_{p}^{*}\psi_{p}) 
\int\prod_{p=A,B}{\cal D}(\varphi_{p}^{*}\varphi_{p}) \times \nonumber\\
& & \exp\left[ A_0+A_{SG}^{(stat)}+A_{K} \right]
\label{e13}
\end{eqnarray}
and
\begin{eqnarray}
A_{K}
&=&\sum_{\sigma}\beta J_{K}\left[\sum_{\omega}\sum_{i_A}
\lambda_{A,-\sigma}\psi_{i_A,\sigma}^{*}(\omega)\varphi_{i_A,\sigma}
(\omega)\right.+\nonumber\\
& & ~~~~~~~~~~~~~~\lambda_{A,\sigma}^{*}\varphi_{i_A,\sigma}^{*}
(\omega)\psi_{i_A,\sigma}(\omega)+\nonumber\\
& &~~~~~~~~~~~~\sum_{\omega}\sum_{j_B}\lambda_{B,-\sigma}\psi_{j_B,\sigma}^{*}
(\omega)\varphi_{j_B,\sigma}(\omega)+\nonumber\\
& & ~~~~~~~~\left.\phantom{\sum_{\omega}}\lambda_{B,\sigma}^{*}\varphi_{j_B,\sigma}^{*}
(\omega)\psi_{j_B,\sigma}(\omega)\right].
\label{e131}
\end{eqnarray}

At this stage, the fluctuations in time and space are explicitly neglected. That
means that in the sum over Matsubara's frequencies, particularly for the spin 
part of action $A_{SG}^{(stat)}$, only the term $\nu=0$ is kept in Eq.\ (\ref{e9}).

The conduction electrons can be integrated in Eq.\ (\ref{e13}) to give
\begin{eqnarray}
\frac{Z^{(stat)}}{Z_0}&=&\int \prod_{p=A,B}{\cal D}(\psi_{p}^{*}\psi_{p})\,
e^{A_{SG}+A_{e\!f\!f}}
\label{e14}
\end{eqnarray}
\noindent where
\begin{eqnarray}
A_{e\!f\!f}=\sum_{i,j}\sum_{\omega,\sigma}
\underline{\Psi}_{i}^{\dagger}(\omega)
\left[\underline{\underline{g}}_{i j}(\omega)\right]^{-1}
\underline{\Psi}_{j}(\omega)
\label{e141}
\end{eqnarray}
with
\begin{eqnarray}
\left[\underline{\underline{g}}_{i j}(\omega)\right]^{-1} = \nonumber
\end{eqnarray}
\begin{eqnarray}
\left[\begin{tabular}{l}
$\left[g_{i j,A}^{0}(\omega)\right]^{-1}-
F_{A}(\omega)$ $~~~~~~~~~~~~~~~~~0~~~~$ \\
$~~~~~~~~~0~~~~~~~~$ $~~~~~~~~\left[g_{i j,B}^{0}(\omega)\right]^{-1}\!-\!
F_{B}(\omega)$
\end{tabular}\right] 
\label{e15}
\end{eqnarray}
\begin{eqnarray}
\underline{\Psi}_{i}^{\dagger}(\omega)&=&\left[\begin{tabular}{cc}
$\psi_{i,A,\sigma}^{*}(\omega)$ $\psi_{i,B,\sigma}^{*}(\omega)$
\end{tabular}\right]\nonumber\\ 
\underline{\Psi}_{j}(\omega)&=&\left[\begin{tabular}{cc}
$\psi_{j,A,\sigma}(\omega)$\\$\psi_{j,B,\sigma}(\omega)$
\end{tabular}\right]
\label{e151}
\end{eqnarray}
\begin{equation}
F_{p}(\omega)=\beta^2 J_{k}^{2}|\lambda_{p}|^{2}\sum_{k}e^{i\vec{k}(\vec{R}_{i}-\vec{R}_{j})}\gamma_{k}(\omega)
\label{e152}
\end{equation}
where $\left[g_{i j,p}^{0}(\omega)\right]^{-1}$ is given in Eq. (\ref{e81})
\noindent and
\begin{eqnarray}
\gamma_{k}(\omega)=\frac{1}{(i\omega-\beta\varepsilon_0)-
\beta\epsilon_k}.
\label{e16}
\end{eqnarray}

The free energy is given by the replica method
\begin{eqnarray}
\beta F &=& \beta J_K (|\lambda_A|^2+|\lambda_B|^2)-\nonumber\\
        & & \lim_{n\rightarrow 0}
\frac{1}{2Nn}\left[\left<\left<Z^{(stat.)}(J_{i j})\right>\right>_{_{CA}}-1\right]
\label{e18}
\end{eqnarray}
\noindent where the configurational average 
$\left<\left<\ ...\ \right>\right>_{_{CA}}$ is performed with the
gaussian distribution given by Eq. (\ref{e2}) applied in Eq.\ (\ref{e14}) which
gives
\begin{eqnarray}
\left<\left<Z^{(stat.)}(J_{i j})\right>\right>_{_{CA}}=
\int D(\psi^{*}\psi)\,e^{A_{e\!f\!f}}\left<\left<Z_{_{SG}}\right>\right>
\label{e19}
\end{eqnarray}
\noindent where
\begin{eqnarray}
\left<\left<Z_{_{SG}}\right>\right>_{CA}=
\prod_{i,j}\left<\left<\exp\left(\beta J_{i j}
\sum_{\alpha}S_{i,A,\alpha}^{z} S_{j,B,\alpha}^{z}\right)\right>\right>_{CA}.
\label{e19n}
\end{eqnarray}

Therefore, the resulting averaged partition function can be linearized by using
the usual Hubbard-Stratonovich transformation and introducting auxiliary fields
of the spin glass part \cite{Korenblit}. The details of this procedure are given in the Appendix.

The free energy can be found using the averaged partition function (see Appendix)
in Eq. (\ref{e18}). Thus, one gets,
\begin{eqnarray}
\beta F&=& 
\beta J_K(|\lambda_A|^2+|\lambda_B|^2)- \beta^2\frac{J^2}{2} q_A q_B
 \nonumber\\&+&\beta^2\frac{J^2}{2} (\tilde{q}_A \tilde{q}_B)
-\beta J_0 m_A m_B
\nonumber\\&-&\lim_{n\rightarrow 0}\frac{1}{2Nn}\int_{-\infty}^{\infty}\prod_{i=1}^{N}D\xi_{i,A}
D\xi_{i,B}
\prod_{\alpha}\int_{-\infty}^{\infty}D z_{i,A}^{\alpha}  D z_{j,B}^{\alpha}\nonumber\\ &\times& 
\exp\left[\sum_{w,\sigma}\ln \det \left[\underline{\underline{G}}_{ij}(\omega|h^{\alpha}_{i,p})\right]^{-1}\right]
\label{e25}
\end{eqnarray}
\noindent where the $\left[\underline{\underline{G}}_{i j}(\omega|h^{\alpha}_{i,p})\right]^{-1}$
in Eq.\ (\ref{e25}) is given by
\begin{eqnarray}
\left[\underline{\underline{G}}_{i j}(\omega|h^{\alpha}_{i,p})\right]^{-1} = \nonumber
\end{eqnarray}
\begin{eqnarray}
\left[\begin{tabular}{l}$(i\omega\!-\!\beta\epsilon_0^{f}\!-\!\sigma
h_{i,A}^{\alpha})\delta_{i,j}\!-\! F_{A}(\omega)$~~~~~~~0~~~~~~ \\
~~~~~~~0~~~~~$~~(i\omega\!-\!\beta\epsilon_0^{f}\!-\!\sigma
h_{i,B}^{\alpha})\delta_{i,j}\!-\!
F_{B}(\omega)$
\end{tabular}\right]
\label{e26}
\end{eqnarray}
\noindent with $F_{p}$  defined in Eq. (\ref{e152}) and 
$h_{i,A}^{\alpha}$ and $h_{i,B}^{\alpha}$ being random gaussian fields (see Eq. (\ref{neweq}) from Apendix)
applied to the sites of one sublattice ($A$ or $B$) which depends on the
parameters of the other sub-lattice. At this point, the static susceptibility $\chi_p$ can be introducted 
related with replica symmetry diagonal order parameter 
$\tilde{q}_{p}=\bar{\chi}_p+q_{p}$ where $\bar{\chi}_p=\chi_p/\beta$. 

The central issue here is to adopt the proper decoupling
approximation which allows one to calculate the matrix $\left[
\underline{\underline{G}}_{i j}(\omega)\right]^{-1}$ \cite{Alba,Magal}. The 
elements given in Eq.\ (\ref{e26}) are referred to the original sublattice
$A\ (B)$ where to each site $i$ there is a random gaussian field 
$h_{i,A}^{\alpha}\ (h_{i,B}^{\alpha})$ applied. The decoupling procedure is to
consider these random fields $h_{i,A}^{\alpha}\ (h_{i,B}^{\alpha})$ as applied
in two fictitious Kondo sub-lattices. Therefore, in each site $\mu$ of
the new sublattice, the applied field $h_{i,A}^{\alpha}\ (h_{i,B}^{\alpha})$
is constant. That is equivalent to replace the Green's function $\left[
\underline{\underline{G}}_{i j}(\omega|h_{i,p}^{\alpha})\right]^{-1}$ by
$\left[\underline{\underline{\Gamma}}_{\mu \nu}(\omega|h_{i,p}^\alpha)\right]^{-1}$  
in  Eq.\ (\ref{e26}). It is, therefore, possible to go to the reciprocal space,
where it is assumed a constant band $\rho (\varepsilon)=\frac{1}{2D}$ for
$-D<\varepsilon<D$ for the conduction electrons in each sublattice. The sum over
the Matzubara
frequencies in Eq.\ (\ref{e25}) can be done following a standard
procedure. Thus, the free energy is found to be
\begin{eqnarray}
\beta F&=& \beta J_K(|\lambda_A|^2+|\lambda_B|^2) + \frac{\beta^2 J^2}{2} \bar{\chi}_A \bar{\chi}_B
+\frac{\beta^2 J^2}{2} (\bar{\chi}_{A}q_B+\nonumber\\ 
& &\bar{\chi}_B q_A)-\frac{\beta J_0}{2} m_A m_B-
\frac{1}{2} \int_{-\infty}^{\infty}\!\!D\xi_{i_A}\!
\int_{-\infty}^{\infty}\!\!D\xi_{j_B}\!\times\nonumber \\
& &\ln\left[\prod_{p=A,B}
\int_{-\infty}^{\infty}D z_{p}e^{E(h_{p})}\right]
\label{e28}
\end{eqnarray}
\noindent with
\begin{eqnarray}
E(h_{p})&=&\frac{1}{\beta D} 
\int_{-\beta D}^{\beta D}dx \left[\cosh\left(\frac{x+h_{p}}{2}\right)+\right.
\nonumber\\
& &\left.\cosh\sqrt{\left(\frac{x-h_{p}}{2}\right)^2+\beta^2 J_k^2 |\lambda_p|^2}~~\right].
\label{e281}
\end{eqnarray}

The saddle point equations for the order parameters $q_p$, $m_p$, $\bar{\chi}_{p}$
and $|\lambda_p|$ \cite{Alba} follow from Eqs.\ (\ref{e28}) and (\ref{e281}).

\section{Results}

The numerical solutions of equations for the order parameters 
$q_p$, $\bar{\chi}_p$, $m_p$ ($p=A,B$) and $|\lambda_p|$ allow one to build up a phase 
diagram of temperature $T$  {\em versus} the set of relevant energy scales 
$J_0$, $J$ and $J_k$ defined in Section II. The thermodynamic phases are 
identified as:(a) an antiferromagnetic (AF) phase corresponding to $m_A=-m_B$ or 
$m_3=0$ (see Eq.\ \ref{e22})-(\ref{e231})); (b) a spin glass phase for 
${q_A}\neq 0$ and $q_B\neq 0$ ($q_3\neq 0$); (c) a Kondo state given by 
$|\lambda_A|\neq 0$ and $|\lambda_B|\neq 0$.

\begin{figure}[t]
\resizebox{.48\textwidth}{!}{
\includegraphics{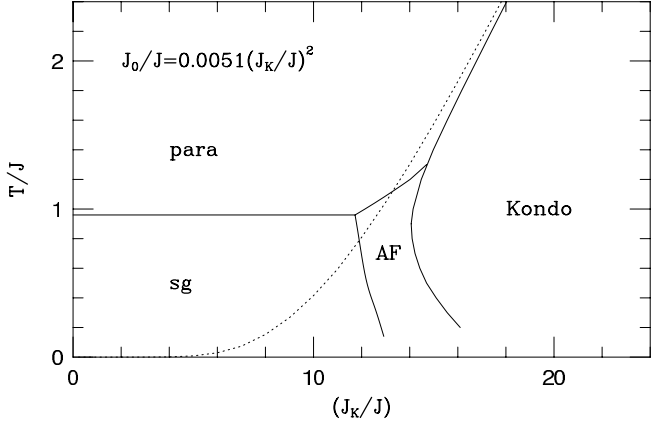}}
\caption{The phase diagram $T/J$ versus $J_{k}/J$ showing the sequences 
of phases SG (Spin Glass), AF (Antiferromagnetism) and a Kondo state. 
The variance $J$ is kept constant. The dotted line means the "pure" Kondo 
temperature.}
\label{fig1} 
\end{figure}

The phase diagrams obtained by the present calculations are derived for several sets of 
parameters $J_{K}/J$ and $J_{0}/J$ considered firstly as independent from each other.
But, as it was discussed in the introduction, the intrasite and intersite exchange terms cannot
be considered as completly independent from each other and a $J_{K}^{2}$-dependence of 
$J_{0}$ was introduced in order to mimic the RKKY interaction \cite{Iglesias}. Thus, a phase 
diagram is shown in Fig.\ (\ref{fig1}) by keeping $J$ constant, taking then:
\begin{equation}
\frac{J_0}{J}=\alpha \left(\frac{J_{k}}{J}\right)^{2}.
\label{e39}
\end{equation}
and by choosing finally  $\alpha=0.0051$. The factor $\alpha$ was 
chosen here to have the AF phase  starting at $J_{k}/J\approx 12$.   

The numerical solutions of the order parameters yield the 
sequence of phases SG-AF-Kondo shown in Fig.\ (\ref{fig1}). The Kondo temperature 
$T_k$ is decreasing for decreasing values of ${J_k}/J$ down to ${J_{k}^*}/J\approx{15}$. 
>From this point onwards there is a phase transition leading to antiferromagnetic 
order with the Neel temperature $T_{N}$ monotonically decreasing with ${J_k}/J$. For 
still smaller values of ${J_k}/J$ (and therefore smaller values of ${J_0}/J$) 
the spin glass behaviour becomes dominant. 

Thus, in this mean field theory it is possible to identify three sorts 
of distinct regimes depending on the strength of the Kondo coupling $J_{k}$. 
The first one (for high $J_{k}$) is the complete screening of the localized magnetic moments due to the 
Kondo effect, which is caracterized by the order parameter 
$|\lambda_p|$ related to the 
formation of d-f singlet throughout the whole two sublattices. In the second one (for decreasing $J_{k}$) , 
the magnetic moments of one particular sublattice, for instance sublattice $A$, 
survive to the screening process (the complete unscreening means $|\lambda_A| = 0$ and 
$|\lambda_B| = 0$) and 
start to be in an antiparallel alignment with an internal field $h_{A}^{\alpha}$, 
which depends on the magnetization and susceptibility of the sublattice $B$, until 
this internal field is spread out to the entire sublattice producing the AF order.
In the last regime ( for $J_k/J < 12$ or $J_0< 0.734J$),  the effects of 
randomness begin to be dominant. In the effective field 
$h_{p}^{\alpha}$ ($p=A,B$), the replica spin glass order parameter $q_{p^{'}}$
($p^{'}=B,A$) component 
starts to be non-null indicating the non-trivial ergodicity breaking leading to a 
spin glass phase at the transition temperature $T_{f}$. Finally, let us remark that we never obtain
theoretically mixed phases where there is the coexistence of the Kondo phase with the other phases. This is 
due to the mean field approximation used here, as already observed \cite{Coqblin,Alba,Magal}.
      
The experimental phase diagram of the alloy $Ce_{2}$$Au_{1-x}$ $Co_{x}$$Si_{3}$ 
\cite{Majundar} can be addressed if the $J_k$ coupling is associated with the 
content of $Co$. The obtained result shown in Fig.\ (\ref{fig1}) displays the 
same sequence of phases at low temperature as the experimental one. In brief, the increasing
$J_{k}$ favours the transition from a SG phase to a AF 
ordering and then to the screening of the localized moments. 
Thus, this 
mean field description is able to account for experimental aspects of 
$Ce_{2} Au_{1-x} Co_{x} Si_{3}$  at low temperatures.  

There is some disagreement related to the location of the  AF line transition. The 
experimental behavior \cite{Majundar} shows the Neel temperature $T_{N}$ 
decreasing apparently towards a quantum critical point (QCP) with increasing 
$Co$ doping. Howewer, recently the spin flipping has been simulated with 
the presence of a transversal field in the x-direction to study a QCP in fermionic 
spin glass \cite{Alba2}. The same approach has been extended to investigate the 
spin glass freezing  in a Kondo lattice \cite{Alba3}. Hence, this method could 
also be used in the present model in order to clarify the role of the QCP in the 
interplay between spin glass and antiferromagnetism in a Kondo lattice.
That will be subject for future investigations.

In conclusion, we have studied in detail the competition between SG, AF and Kondo
phases in a mean field approach of Kondo-lattice disordered alloys. Using the peculiar
relationship (\ref{e39}) between the different parameters of the model, we have obtained 
a phase diagram showing the sequence SG-AF-Kondo phases in good agreement with 
the experimental phase diagram of the disordered alloys $Ce_{2}Au_{1-x}Co_{x}Si_{3}$, 
if we assume that $J_{K}$ increases with increasing Cobalt concentration. Other phase
sequences, involving also SG, AF and Kondo phases, obtained in disordered Uranium alloys, 
are not accounted for, at present, by our model. But the choice of parameters and their 
relationship with the varying concentration in these alloys are really delicate 
questions which are not answered at present. Moreover, in such Uranium systems, the Kondo
phases have clearly a Non-Fermi liquid behavior which is not described here.  Finally, 
the present model can account for the presence of SG, AF and Kondo phases, but further 
work is necessary in order to obtain a better agreement with experimental data for Cerium and Uranium 
disordered alloys.

\noindent {\small {\bf Acknowledgment} The numerical calculations were performed at LANA
(Departamento de Matem\'atica, UFSM) and at LSC (Curso de Ci\^encia da
Computa\c{c}\~ao, UFSM). This work was partially supported by the brazilian
agencies FAPERGS (Funda\c{c}\~ao de Amparo \`a Pesquisa do Rio Grande do Sul) 
 and CNPq (Conselho Nacional de Desenvolvimento Cient\'{\i}fico e
 Tecnol\'ogico). One of us (B.C.) thanks also the CNRS-CNPq french-brazilian cooperation.}

\section*{Appendix}

The averaged partition function can be given by \cite{Korenblit}:
\begin{eqnarray}
\left<\left<Z_{_{SG}}\right>\right>&=&
\int\prod_{p=A,B}\prod_{\alpha}
\sqrt{\frac{N\beta J_0}{2\pi}}dM_{p}^{\alpha}\nonumber\\ 
&\times& \int\prod_{p=A,B}\prod_{\alpha\beta}
\sqrt{\frac{N\beta^2 J^2}
{2\pi}}dQ_{p}^{\alpha\beta}\nonumber\\ 
&\times&
\int\prod_{\alpha\beta}
\sqrt{\frac{N\beta^2 J^2}
{2\pi}}dQ_{3}^{\alpha\beta}
\int\prod_{\alpha}
\sqrt{\frac{N\beta J_0}
{2\pi}}dM_{3}^{\alpha}\nonumber\\
&\times&\exp\left\{-N\left[
-\frac{1}{N}\ln\ \Lambda(M_3^\alpha,Q_{3}^{\alpha\beta})
\right.\right.
\nonumber \\
&+&\frac{\beta^2 J^2}{2}\sum_{\alpha \beta}\left((Q_3^{\alpha\beta})^{2}
+\sum_{p=A,B} (Q_p^{\alpha\beta})^{2}\right) \nonumber \\
&+&
\left.\left.
\frac{\beta J_o}{2}\sum_{\alpha}
\left((M_3^\alpha)^2
+\sum_{p=A,B}(M_p^\alpha)^2\right)
\right]\right\}
\label{e20}
\end{eqnarray}
\noindent where the $\Lambda(M_3^\alpha,Q_{3}^{\alpha\beta})$ is defined as
\begin{eqnarray}
\Lambda&=&\int D(\psi^*\psi)
\exp\left\{ \sum_{\alpha}\sum_{i,j}
\sum_{\omega\sigma}\right.\nonumber\\
&&\underline{\Psi}_{i,\alpha}^{\dagger}(\omega)\left[\underline{\underline{g}}
_{i j}(\omega)\right]^{-1}\underline{\Psi}_{j,\alpha}
(\omega)
\nonumber \\
&+& i\beta J_0 \sum_{\alpha}\left[\sum_{i,A}S_{i,A,\alpha}^{z}+\sum_{j,B}
S_{j,B,\alpha}^{z}\right]M_{3}^{\alpha}\nonumber \\
&+&\beta J_0 \sum_{\alpha}\left[\sum_{i}S_{i,A,\alpha}^{z}M_A^{\alpha}+
\sum_{j}S_{j,B,\alpha}^{z}M_B^{\alpha}\right]
\nonumber \\
&+& 4 i\beta^2 J^2 \sum_{\alpha\beta}\left[ \sum_{i}
(S_{i,A,\alpha}^{z}S_{i,A,\beta}^{z})Q_A^{\alpha\beta}+\right.\nonumber\\
& &~~~~~~~~~~~~~~~\left.\sum_{j}(S_{j,B,\alpha}^{z}S_{j,B,\beta}^{z})
               Q_B^{\alpha\beta}\right]\nonumber \\
&+& 4\beta^2 J^2 \sum_{\alpha\beta}\left[ \sum_{i}
S_{i,A,\alpha}^{z}S_{i,A,\beta}^{z}+\right.\nonumber\\
& &~~~~~~~~~~~~~~\left.\left.\sum_{j}S_{j,B,\alpha}^{z}S_{j,B,\beta}^{z}
                   \right]Q_3^{\alpha\beta}\right\}~.
\label{e21}
\end{eqnarray}

>From Eqs.\ (\ref{e20}) and (\ref{e21}), one finds the saddle point solution for
these auxiliary fields \cite{Korenblit}:
\begin{equation}
M_{3}^{\alpha}=\frac{i}{N}\left< \sum_{i}S_{i,A,\alpha}^{z}+
\sum_{j}S_{j,B,\alpha}^{z}\right>=2i\ m_3^\alpha
\label{e22}
\end{equation}
\begin{equation}
M_{p}^{\alpha}=\frac{1}{N}\left<\sum_{i} S_{i,p,\alpha}^{z}\right>= m_p^\alpha
; ~~~~~p=A,~B
\end{equation}
\begin{equation}
Q_{3}^{\alpha\beta}=\frac{4}{N}\left< \sum_{i}S_{i,A,\alpha}^{z} S_{i,A,\beta}^{z}+
\sum_{j}S_{j,B,\alpha}^{z} S_{j,B,\beta}^{z}\right>=2 q_3^{\alpha\beta}
\label{e23}
\end{equation}
\begin{equation}
Q_{p}^{\alpha\beta}=\frac{4i}{N}\ \left<\sum_{i} S_{i,p,\alpha}^{z} S_{i,p,\beta}^{z}\right>
=i q_p^{\alpha\beta}; ~~~~~p=A,~B
\label{e231}
\end{equation}

It has been assumed, within the replica symmetry ansatz for the auxiliary fields (see
Eqs.\ (\ref{e22})-(\ref{e231})), that $q_{3}^{\alpha\beta}=q_3$,
$q_{p}^{\alpha\beta}=q_{p}$ and $q_{p}^{\alpha\alpha}=\tilde{q}_{p}$
(analogously $m_3^\alpha=m_3$ and $m_{p}^\alpha =m_{p}$). The sum over
replica index produces again quadratic forms which can be linearized by new
auxiliary fields. Thus, for $\Lambda(m_3, q_3)$ one has
\begin{eqnarray}
\Lambda&=&\int D(\psi^* \psi) \exp\left\{\sum_{i,j,\omega,\alpha}\underline{\Psi}_{i,\alpha}^{\dagger}(\omega)
\left[\underline{\underline{g}}_{i j}(\omega)\right]^{-1}
\underline{\Psi}_{j,\alpha}(\omega)\right\}\nonumber \\
&\times&
\int_{-\infty}^{\infty}\prod_{i=1}^{N} D \xi_{i,B}\prod_{\alpha}\int_{-\infty}^{\infty}D z_{i,B}^{\alpha}
\exp\left[h^{\alpha}_{i,B}
S_{i,B,\alpha}^{z}\right]
\nonumber\\
&\times&
\int_{-\infty}^{\infty}\prod_{i=1}^{N} D \xi_{i,A}\prod_{\alpha}\int_{-\infty}^{\infty}D z_{i,A}^{\alpha}
\exp\left[h^{\alpha}_{i,A}
S_{i,A,\alpha}^{z}\right]
\label{ee24}
\end{eqnarray}
where the random field $h^{\alpha}_{i,p}$, introduced in the previous equation, is defined as
\begin{eqnarray}
h_{i,p}^{\alpha}=\beta J \left(\sqrt{2q_{p^{'}}}\xi_{i,p}
+\sqrt{2(\tilde{q}_{p^{'}}-q_{p^{'}})}z_{i,p}^{\alpha}\right) \nonumber\\-
\beta  J_0 m_{p^{'}}\,\,\,\,\,(p\neq p^{'})
\label{neweq}
\end{eqnarray}
with $\displaystyle S_{i,p,\alpha}^{z}=\frac{1}{2}
\sum_{\omega,\alpha,\sigma} \sigma(\psi_{i,p,\sigma}^{\alpha})^{*}(\omega)
\psi_{i,p,\sigma}^{\alpha}(\omega)$ and 
$\displaystyle Dx=$\newline$(e^{-x^2/2}/\sqrt{2\pi})\,dx$.

The functional integral in Eq. (\ref{ee24}) can be found following 
standard procedure \cite{Negele}.


\end{document}